\documentclass{PoS}

\title{Melting instantons, domain walls, and large N }

\ShortTitle{Melting instantons}

\author{\speaker{Hank Thacker}\\
          University of Virginia\\
        E-mail: \email{hbt8r@virginia.edu}}

\abstract{Monte Carlo studies of $CP^{N-1}$ sigma models have shown that the
structure of topological charge in these models undergoes a sharp transition at $N=N_c\approx 4$.
For $N<N_c$ topological charge is dominated by small instantons, while for $N>N_c$ it is 
dominated by extended, thin, 1-dimensionally coherent membranes of topological charge, which can be 
interpreted as domain walls between discrete quasi-stable vacua. These vacua differ by a unit of 
background electric flux. The transition can be identified as the delocalization of topological
charge, or ``instanton melting,'' a phenomenon
first suggested by Witten to resolve the conflict between instantons and large $N$ behavior.
Implications for $QCD$ are discussed. 
         }

\FullConference{The XXVI International Symposium on Lattice Field Theory \\
		 July 14 - 19, 2008\\
		 Williamsburg, Virginia, USA}

\begin{document}

\section{Introduction}

The structure of topological charge fluctuations in the QCD vacuum is a central issue
in the study of low energy hadron dynamics. The role of topological charge in resolving
the axial U(1) problem and providing a mass to the $\eta'$ meson is well known and
has been extensively studied in lattice calculations \cite{Bardeen}. In addition to resolving the $U(1)$
problem, it is likely that topological charge fluctuations are responsible for spontaneous
chiral symmetry breaking and the appearance of a quark condensate in the QCD vacuum. 
For example, in the context of the instanton liquid 
model, the near-zero Dirac eigenmodes needed to form a quark condensate are provided by
the approximate t'Hooft zero modes of the instantons and anti-instantons \cite{Diakonov}.
Although one may question the assumptions of the instanton liquid 
model, it can be argued that the connection between topological charge and chiral
symmetry breaking is more general and does not depend on the specifics of the instanton model. 
Any sufficiently strong coherent
regions of positive (negative) topological charge will attract 
left (right)- handed quarks and produce near-zero Dirac eigenmodes
localized in the region of the topological charge fluctuation. As I will 
discuss here, a combination of theoretical arguments \cite{Witten79,Witten98} 
and lattice studies \cite{Horvath03,Ilgenfritz} point toward a new paradigm for the QCD
vacuum, in which coherent topological charge structure 
comes not in the form of instantons, but of extended, thin, coherent
membranes of codimension one. In a large-N chiral Lagrangian framework, these membranes
appear as domain walls between discrete chiral phases \cite{Witten79}. From the point of
view of AdS/CFT string/gauge duality, these topological charge membranes
can be interpreted as the holographic image of D6-branes (wrapped around an $S_4$),
which arise naturally as carriers of Ramond-Ramond (RR) charge in IIA string theory \cite{Witten98}.
The $\theta$ parameter in QCD arises from a nonzero Wilson line of the RR potential 
around a compact $S_1$, and the corresponding domain wall separates vacua with $\theta$
parameters differing by $\pm 2\pi$. In the string theory, the statement that $\theta$ steps by
an integer multiple of $2\pi$ across the membrane corresponds to the
quantization of RR charge on the D6-brane. It is interesting that the gauge theory construct
which corresponds holographically to the string theory D6-brane was introduced long ago by
Luscher \cite{Luscher78}. This ``Wilson bag'' operator is the integral over a 3-surface of
the 3-index Chern-Simons tensor of Yang-Mills theory (the operator whose curl is the topological charge). In this
talk I will make considerable use of the analogy between 4-dimensional Yang-Mills theory and 
2-dimensional $CP^{N-1}$ sigma models. In the 2D models, the Wilson bag operator reduces to the ordinary
Wilson loop (which in 2 dimensions is also an integral of the Chern-Simons flux 
$j_{\mu}^{CS}\equiv \varepsilon_{\mu\nu}A_{\nu}$ over a codimension-one surface) \cite{Luscher78}.

\begin{figure}
   \begin{center}
     \vskip -0.15in
     \centerline{
     \includegraphics[width=9.5truecm,angle=-90]{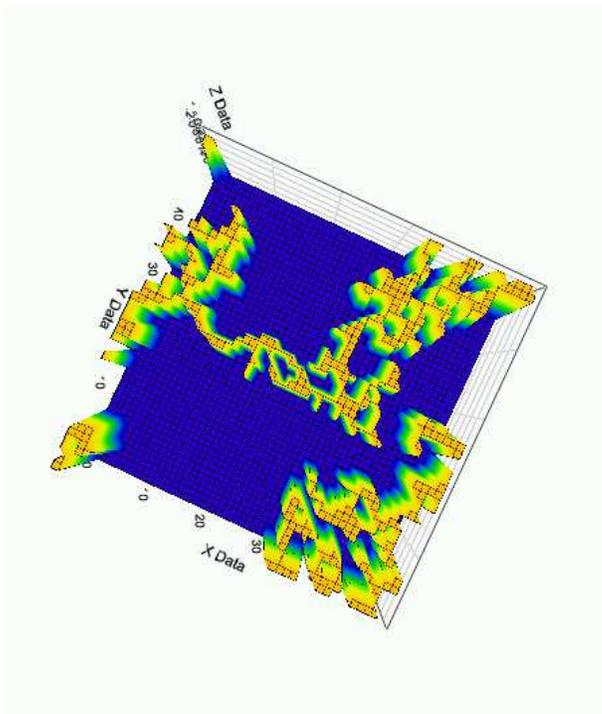}
     }
     \vskip 0.15in
     \caption{A coherent structure in the topological charge distribution of a typical
Monte Carlo configuration for $CP^5$ at $\beta=1.2$.}
     \label{fig:1Dstructure}
   \end{center}
\end{figure}

In the large N limit of an asymptotically free gauge theory, we expect that instantons will disappear
in favor of codimension one topologically charged domain walls separating discrete, quasi-stable
vacua. Monte Carlo studies have borne this out for both
4-dimensional $SU(3)$ gauge theory \cite{Horvath03,Ilgenfritz} and 2-dimensional $CP^{N-1}$ models 
\cite{Ahmad,Lian} with $N>4$. In both these cases, the Monte Carlo results are consistent with expectations 
based on the large $N$ approximation. But the $CP^{N-1}$ models 
have the interesting feature that the topological 
charge structure undergoes a rather sharp transition as a function of $N$ at around $N_c\approx 4$,
which can be identified as the instanton melting point. For $N>4$ the gauge configurations are dominated
by codimension one membranes, but below this value of N, i.e. for $CP^1$ and $CP^2$,
the topological fluctuations are dominated by small instantons \cite{Lian,Luscher82}. 
The $CP^{N-1}$ models thus provide an opportunity to study the transition from instanton dominance to 
domain wall dominance and to identify the nature of the instability that leads to instanton
melting. [Note: The small instantons that dominate $CP^1$ and $CP^2$ have radii of order the lattice spacing
and lead to a non-scaling contribution to the topological susceptibility, as discussed by Luscher \cite{Luscher82}.
For $CP^2$ it is possible to define a lattice version of the model which eliminates
the small instantons \cite{Petcher83}. But $CP^1$
will be instanton dominated for any lattice action and has a divergent $\chi_t$ in the scaling limit.
For $N\geq 4$, $\chi_t$ is found to scale properly to a finite value in the continuum limit \cite{Lian}.] 

The $CP^{N-1}$ models are asymptotically
free, classically conformally invariant, and acquire a mass scale via a conformal anomaly (dimensional transmutation),
just as in QCD. Most importantly for this discussion, they have classical instanton solutions 
whose contributions are exponentially suppressed at large N, like QCD instantons. As emphasized in Witten's
original discussion \cite{Witten79}, this exponential suppression is an indication that, at sufficiently
large $N$, instantons melt or disappear from the path integral 
in favor of other topological charge fluctuations associated
with the confining vacuum which are only suppressed by $1/N$. After considering the instanton melting phenomenon
in the $CP^{N-1}$ models, I will then discuss the implications for 4-dimensional QCD.

\section{Topological charge structures in $CP^{N-1}$ and $QCD$}

\begin{figure}
   \begin{center}
     \vskip -0.15in
     \centerline{
     \includegraphics[width=9.5truecm,angle=0]{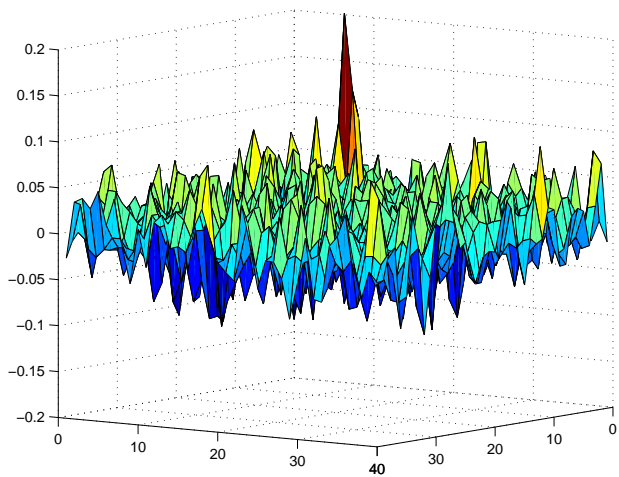}
     }
     \vskip 0.15in
     \caption{Topological distribution for a $CP^2$ configuration with total Q=+1. The prominent instanton
peak seen contains approximately one unit of topological charge.}
     \label{fig:instanton}
   \end{center}
\end{figure}

The discovery of 3-dimensional topological charge membranes in 4D $SU(3)$ QCD and of coherent
1-dimensional structures in 2-dimensional $CP^{N-1}$ gauge configurations \cite{Ahmad} provides clear evidence of long
range topological charge structure. The $CP^{N-1}$ results also strongly support the interpretation of this
structure in terms of discrete $\theta$ vacua and domain walls. To see the emergence of 1-dimensional structure,
it is instructive to plot the largest connected coherent 
structure in a typical gauge configuration. (Here nearest-neighbor sites are connected if they have the same sign
of $q$.) Comparing with similar plots for randomly generated $q(x)$ distributions, the emergence of 1-dimensional
structures is clear. Fig. \ref{fig:1Dstructure} shows the largest structure in a typical configuration for 
$CP^5$ on a $50\times 50$ lattice at $\beta=1.2$ (correlation length $\approx 20$). Nearly all of the gauge
configurations have largest structures which extend over the length of the lattice in some direction. 
The 1-dimensional coherence of the observed topological charge structures
clearly extends well beyond the physical correlation length of the system. This is even more clear in the
$QCD$ studies \cite{Horvath03}, where the correlation length determined by the pseudoscalar glueball mass is
very short ($<.1 fm$) compared to the observed coherence of the topological charge sheets, which extend 
over the whole lattice. It is also worth remarking that the observed structure cannot be dismissed as an 
artifact of the overlap construction of $q(x)$. In the $CP^{N-1}$ case, 
the very same 1-dimensional structures can also be identified in
the distribution obtained from the ultralocal $q(x)$ operator constructed from the 
log of the plaquette after using a simple 1-hit nearest-neighbor
smoothing procedure. In fact, in general for the 2D models, 
the overlap topological charge at a site is reasonably well approximated
by the average of the four plaquette charges around the site. 

For $CP^1$ and $CP^2$ the topological charge structure is qualitatively different than for $CP^3$ and higher.
It is dominated by small instantons which typically have a radius of one or two lattice spacings.
FIg. \ref{fig:instanton} shows the overlap topological charge distribution for a typical $CP^2$ configuration
at $\beta=1.8$. This configuration has a global topological charge of $+1$. The prominent peak seen is a
small instanton, containing approximately one unit of TC. The rest of the distribution integrates to approximately
zero. At sufficiently large $\beta$ these small instantons totally dominate the TC distribution for $CP^1$ and $CP^2$. 
In the course of a Monte Carlo run, the tunneling of the global charge from one integer to another is invariably 
accompanied by the appearance or disappearance of an instanton or anti-instanton. This is in marked contrast to
the results for $CP^5$ and $CP^9$, which showed no trace of instantons in any configuration. 
 
\section{Theta dependence, discrete vacua, and domain walls}

\begin{figure}
   \begin{center}
     \vskip -0.15in
     \centerline{
     \includegraphics[width=9.5truecm,angle=0]{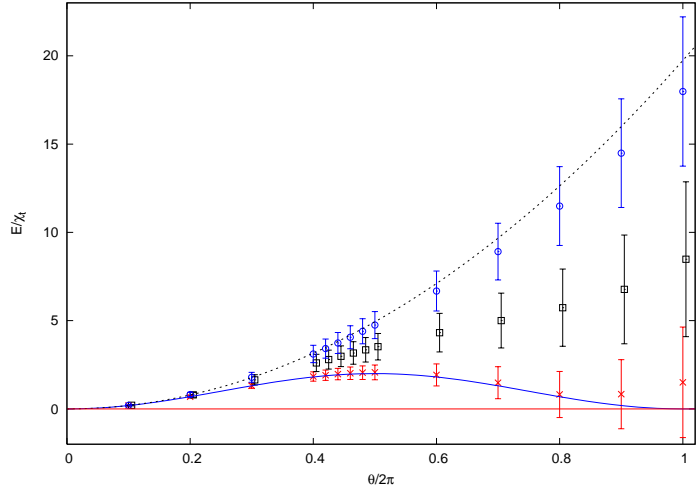}
     }
     \vskip 0.15in
\caption{The free energy density $\varepsilon(\theta)$ for $CP^1$ ($\times$'s), $CP^5$ ($\Box$'s),
and $CP^9$ (o's) extracted from fractionally charged
Wilson loops. The lower and upper curves are the instanton gas
and large $N$ predictions, normalized to the same topological susceptibility. Note that, for $\theta/2\pi>\frac{1}{2}$ the large N curve is
interpreted as the energy of the false (unscreened) vacuum.
}
     \label{fig:gsenergy}
   \end{center}
\end{figure}

Recent results from an analysis of fractionally charged Wilson loops \cite{fracloops} 
in $CP^1$, $CP^5$, and $CP^9$, provide 
further evidence of the transition from an instanton gas to large N behavior. This transition is directly
reflected in the change of behavior of the ground state energy $\varepsilon(\theta)$ in the region $\theta
\approx \pi$. First consider the results for $\theta<\pi$. As shown in Fig. \ref{fig:gsenergy} and \cite{fracloops}, the behavior of $\varepsilon(\theta)$ for 
$CP^1$ is in good agreement with the dilute instanton gas prediction
\begin{equation}
\label{eq:instantongas}
\varepsilon(\theta) = \chi_t(1-\cos\theta)
\end{equation}
while, for $\theta<\pi$, $CP^9$ agrees well with the large $N$ prediction
\begin{eqnarray}
\varepsilon(\theta) & = & \frac{1}{2}\chi_t\theta^2,\;\;\;\;\; \theta<\pi \label{eq:largeN_1} \\
&=&\frac{1}{2}\chi_t(2\pi-\theta)^2 \;\; \pi<\theta<2\pi  \label{eq:largeN_2}
\end{eqnarray}
For $CP^5$ the results lie between the instanton and large $N$ predictions, but both $CP^5$ and $CP^9$ clearly
exhibit a positive, nonvanishing slope at $\theta=\pi$. Because of symmetry around $\pi$,
\begin{equation}
\label{eq:reflection}
\varepsilon(\theta)=\varepsilon(2\pi-\theta)
\end{equation}
this implies that $\varepsilon(\theta)$ has a cusp at $\theta=\pi$, indicating a first order phase transition between
discrete vacua for both $CP^5$ and $CP^9$. On the other hand for $CP^1$, the vanishing slope at 
$\theta=\pi$ appears to rule out a first order transition for this model. (The data near $\theta=\pi$ is not 
accurate enough to observe the second order transition expected for $CP^1$ from Haldane's arguments.)

An even more illuminating distinction between $CP^1$ and the large $N$ models is seen in the behavior of 
the apparent value of $\varepsilon(\theta)$ for $\theta>\pi$. The data plotted in Fig. \ref{fig:gsenergy} are 
the values of $\varepsilon(\theta)$ extracted from the area law for a large Wilson loop of charge $\theta/2\pi$.
For $CP^1$, the results agree with the instanton gas formula (\ref{eq:instantongas}) over the entire range
$0<\theta<2\pi$, and are consistent with the reflection symmetry (\ref{eq:reflection}). In particular,
$\varepsilon(\theta)$ returns to zero and the confining potential vanishes at $\theta=2\pi$. This shows
that an integer charged loop is completely screened. But for $CP^5$ and $CP^9$, the measured $\varepsilon(\theta)$ 
violates the reflection symmetry and continues to increase beyond $\theta=\pi$. In fact, for $CP^9$, it continues
to follow the extrapolated large $N$ formula (\ref{eq:largeN_1}) rather than (\ref{eq:largeN_2}). In the large $N$ framework, there are two possible
discrete vacua in the interior of the Wilson loop. One has a background electric flux of $\theta/2\pi$ (unbroken
string), the other has a flux of $\theta/2\pi-1$ (broken string). For $\theta>\pi$ the broken string is the true
(screened) vacuum. However, the Wilson line has a much larger overlap with the unbroken string. (A similar phenomenon
occurs in full QCD with dynamical quarks, where string breaking should occur but is difficult to observe in Wilson loops.)
The results for $CP^5$ and $CP^9$ at $\theta>\pi$ clearly demonstrate the existence of a discrete, 
quasi-stable false vacuum state which differs from the true vacuum by one unit of electric flux.
By contrast, the $CP^1$ model shows no evidence of a quasi-stable electric flux string. The Wilson loop
area law simply vanishes for an integer charged loop. For fractional charge, the fluctuating number of instantons
inside the loop generates a confining potential by randomizing the phase of the loop. But when small instantons
dominate, the amount of topological charge inside the loop is typically close to an integer, so the phase of an 
integer charged loop is not randomized. 

\section{Implications for QCD}

The behavior of the $CP^{N-1}$ models suggests a heuristic description of the instanton melting transition:
Above some value of $N$ the action begins to favor large instantons over small ones. But the tendency for small
instantons to grow into large ones is in conflict with a fundamental requirement on the two-point topological
charge correlator, i.e. that it has to be negative for nonzero separation in the continuum theory\cite{Seiler}. (Instantons
with radius of order lattice spacing only contribute to the positive delta-function contact term in the correlator
and thus do not violate the negativity requirement.)
As an instanton grows in radius the negativity of the correlator forces it to become a thin hollow
bubble (a Wilson loop excitation in 2D, or a Wilson bag in 4D). The Wilson bag is screened by an anti-bag which 
appears inside the bag (the analog of string-breaking in 2D). Further growth produces an 
alternating sequence of concentric bags. The original unit
of topological charge has delocalized and the vacuum is filled with alternating-sign, codimension one membranes,
in agreement with the observed structure of Monte Carlo generated gauge configurations.  
In an asymptotically free theory, a weak coupling calculation of the instanton contribution to the path integral
is given in terms of an integral over instanton radius of the form $\int d\rho \rho^{\alpha(N)}$. A semiclassical
estimate of (actually a lower bound on) the instanton melting point $N_c$ is given by the ``tipping point'' of
the integral over radius, where $\alpha(N_c)= -1$. For $N>N_c$ the integral diverges for large instantons and 
instantons are presumed to melt, as described. This gives a melting point of $N_c=2$ for $CP^{N-1}$ and
$N_c=12/11$ for 4D gauge theory. This along with the dominance of codimension one structures observed in Monte Carlo 
configurations suggests that 4-dimensional $SU(3)$ gauge theory is in the large $N$ phase. 

The view of the QCD vacuum as a laminated ``topological sandwich'' of alternating sign membranes has some appealing
features. A nonzero value of the $\theta$ parameter can be thought of as analogous to 
an electric field transverse to the membranes.
The fact that the topological susceptibility is nonzero results from the fact that the topological sandwich vacuum
is a polarizable medium. The $U_A(1)$ problem is resolved ala Witten-Veneziano. A massless Goldstone pion is naturally
interpreted as a quark-antiquark pair propagating along adjacent, oppositely charged TC membranes via delocalized surface
modes on these membranes. Finally, it can be plausibly argued that the topological sandwich vacuum  leads to confinement.
In the $CP^{N-1}$ models, charge confinement occurs by quite different mechanisms for the small-$N$ instanton vacuum
and the large-N topological sandwich vacuum. In the instanton models, although fractional charge is confined, there is
no confining force between integer charges. However, for large $N$, the effect of the topological charge membranes is to produce a 
quasi-stable electric flux string and a linear potential, even between integer charges. As is well-known, the instanton
confinement mechanism does not generalize to four dimensions. On the other hand, 
the large-$N$ confinement mechanism induced by
codimension one membranes (which disorder a Wilson loop and lead to an area law) seems quite likely to generalize
to four-dimensional gauge theory.


\begin{thebibliography}{99}

\bibitem{Bardeen} W. Bardeen, et al., Phys. Rev. D62:114505 (2000); Phys. Rev. D65:014509 (2001); Phys. Rev. D69:054502 (2004); Phys. Rev. D70:117502 (2004).

\bibitem{Diakonov} D. Diakonov and V. Petrov, Phys. Lett. B147:351 (1984);

\bibitem{Witten79} E. Witten, Nucl. Phys. B149, 285 (1979).

\bibitem{Witten98} E. Witten, Phys.~Rev.~Lett. 81: 2862 (1998).

\bibitem{Horvath03} I. Horvath et al, Phys. Rev. D68, 114505 (2003).

\bibitem{Ilgenfritz} E.-M.~Ilgenfritz, et. al, Phys. Rev. D76:034506 (2007).

\bibitem{Luscher78} M.~Luscher, Phys. Lett. B78:465 (1978).

\bibitem{Ahmad} S.~Ahmad, J.~Lenaghan, and H.~Thacker, Phys. Rev. D72:114511 (2005).

\bibitem{Lian} Y.~Lian and H.~Thacker, Phys. Rev. D75:065031 (2007).

\bibitem{Luscher82} M.~Luscher, Nucl. Phys. B200:61 (1982).

\bibitem{Petcher83} D.~Petcher and M.~Luscher, Nucl. Phys. B225:53 (1983).

\bibitem{fracloops} P.~Keith-Hynes and H.~Thacker, Phys. Rev. D78:025009 (2008).

\bibitem{Seiler} E.~Seiler and I.~O.~Stamatescu, MPI-PAE/Pth 10/87.

\end{thebibliography}
\end{document}